\documentstyle[aps,epsf]{revtex}
\draft
\begin{document}
\title{Dark States and Interferences in Cascade Transitions of Ultra-Cold Atoms in
a Cavity}
\author{R. Arun and G. S. Agarwal}
\address{Physical Research Laboratory, Navrangpura, Ahmedabad 380 009,
India}
\date{\today}
\maketitle
\begin{abstract}
We examine the competition among one- and two-photon processes in an ultra-cold,
three-level atom undergoing cascade transitions as a result of its interaction 
with a bimodal cavity.  We show parameter domains where two-photon transitions
are dominant and also study the effect of two-photon emission on the mazer
action in the cavity. The two-photon emission leads to the loss of detailed
balance and therefore we obtain the photon statistics of the cavity field 
by the numerical integration of the master equation. The photon distribution 
in each cavity mode exhibits sub- and super- Poissonian behaviors depending on
the strength of atom-field coupling. The photon distribution becomes identical
to a Poisson distribution when the atom-field coupling strengths of the modes 
are equal.   
\end{abstract}
\newpage
\section{introduction}
The dynamics of atom-field coupling in high quality cavities has been
an interesting tool to verify the predictions of quantum electrodynamics (QED)
on the radiation-matter interaction \cite{{berman},{eberly},{spont},{rabi},{rmp}}. 
The properties of atoms in cavities can be modified in a controlled way by the proper
design of cavity geometry and its quality factors. For example, the spontaneous
emission of an atom can be enhanced or inhibited inside a cavity \cite{spont}.
Cavities with high quality factors are especially attractive for studying the
regime of strong interaction between the atoms and the quantized fields. 
Interesting effects of strong interaction such as collapse and revival of 
Rabi oscillations, vacuum Rabi splitting, atom-atom and atom-field entanglements
have been predicted and observed \cite{{rabi},{rmp}}. The generation of atom-field
entangled states in cavity QED finds many applications in quantum information
processing, logic gates etc. A very important result of strong interaction was
the micromaser where the excited atoms go successively through a high quality 
cavity in a time much smaller than the characteristic decay times. This leads
to build up of a steady state field in the cavity which has properties quite
different from laser fields. The theoretical works on micromaser 
\cite{{meystre},{lugiato}} suggested the sub-Poissonian photon statistics
which is purely due to the quantum nature of the field. The operation of the 
micromaser has been realized \cite{walther} and Weidinger et al verified the 
existence of trapping states in the micromaser \cite{weidinger}. The two-photon
micromaser as well as the microlaser were also realized \cite{{brune},{ann}}.
The theory of two-mode micromaser operating on three-level systems has been 
developed \cite{fam}.
 
In all these works, the external motion of the atom is treated classically
while its interaction with the field is quantum mechanical. By considering 
ultra-cold atoms, Scully et al \cite{scully} discovered that the
quantum treatment of center-of-mass (c.m.) motion of an atom leads to a
completely new kind of induced emission in a cavity. The dressed state  
analysis shows that the cavity field can act like a quantum mechanical potential         
for an ultra-cold, incident atom. The nature of the potentials depends on 
the mode profile of the cavity as well as the atom-field coupling strengths
\cite{{rempe},{solano}}. Thus the atom besides changing its electronic  
states can be either reflected or transmitted through the cavity. This reflection
or transmission of the atoms is very similar to those of a particle interacting
with potential barriers or wells. We have recently shown that tunneling of an
ultra-cold, two-level atom in the excited state through two successive cavities
exhibits transmission resonances similar to those of an electron tunneling through 
semiconductor double barriers \cite{gsa}. The reflection and transmission through
cavity induced potentials give rise to many interesting features in the photon 
statistics and the spectrum of the micromaser pumped by ultra-cold atoms 
\cite{meyer}. Further, the works on quantum treatment of the atomic motion have     
been extended to study the interaction of two- and three- level atoms with a 
single mode field \cite{zhang}. Considering the quantized motion of $\Lambda$-type
three-level atoms, the photon statistics of a two-mode micromaser was discussed
\cite{arun}. In the $\Lambda$-type scheme, an excited atom can
make only one photon transition to either of the two ground levels. In this paper,
we examine the photon statistics of a two-mode micromaser pumped by ultra-cold,
cascade three-level atoms. The excited atom in the cascade transition emits
photons sequentially in each of the two cavity modes. Thus, the field mode which is
resonant with the upper transition of the $\Xi$-type atoms, gets populated by both
one and two-photon transitions from the excited state. We discuss the effect of
two-photon emissions from the excited atoms on the photon statistics of the 
micromaser. We make use of methods similar to those in Ref. \cite{meyer}.
  
The organization of the paper is as follows. In Sec. II, we describe  
the interaction of ultra-cold, $\Xi$-type three-level atoms with the bimodal  
field in the cavity. We show that the field can always induce two-photon transition  
in an excited atom when the corresponding one-photon transition is forbidden. 
In Sec. III, we study the mazer action and derive the master equation for the 
reduced density operator of the field in the cavity. In Sec. IV, the numerical 
results are presented for the steady state photon distribution in each mode of
the bimodal field. We find that the photon distribution in each mode exhibits 
Poissonian statistics if the atom-field coupling strengths of the cavity modes 
are equal and when the two-photon transitions from the excited atoms are 
dominant. For different coupling strengths of the cavity modes, the dominant     
two-photon effects in transitions can also lead to sub- and super-Poissonian 
statistics of photons in the steady state field. Finally, we consider the mazer
action when the pumping atoms have kinetic energies close to the vacuum coupling 
energy. In this case, one-photon effects also become important and these lead
to different photon statistics for the two cavity modes.  
   
\section{one and two photon processes in ultracold atoms in a cavity}
We consider a bimodal cavity of length $L$ pumped steadily by a beam of ultra-cold 
three-level atoms in the cascade configuration. The scheme of our model
is shown in Fig. 1. The transitions $a \rightarrow b_1$ and $b_1 
\rightarrow b_2$ are dipole allowed while the direct transition $a
\rightarrow b_2$ is dipole forbidden. Thus the atom in the excited 
level $a$ can reach the ground level $b_2$ only through the two-photon
transition $a \rightarrow b_1 \rightarrow b_2$. The frequencies of the
two cavity modes 1 and 2 are tuned to those of atomic transitions $a
\rightarrow b_1$ and $b_1 \rightarrow b_2$ respectively. The Hamiltonian
describing this resonant atom-field interaction including the quantized 
motion of center-of-mass of the atoms along the $z$ direction is given by 
\begin{equation}
H = H_A + H_F + H_{AF}~,
\label{ham1}
\end{equation}
where $H_A (H_F)$ is the Hamiltonian of the free atom (field) and $H_{AF}$
is the interaction Hamiltonian describing the atom-field interaction in the
dipole and the rotating wave approximations: 
\[
H_A = \frac{p^{2}_z}{2m} + \hbar{\Omega}_a|a\rangle\langle a| +
        \sum_{\alpha = 1}^{2} \hbar{\Omega}_{b_{\alpha}} |b_{\alpha}\
        \rangle\langle b_{\alpha}| \nonumber~,
\]

\[
H_F = \sum_{\alpha =1}^{2} \hbar{\omega}_{\alpha} a_{\alpha}^{\dag}
        a_{\alpha}~,
\]

\begin{equation}
H_{AF} = \hbar g_{1}(a_1~ |a\rangle\langle b_1| + |b_1\rangle\langle a|
~a_1^{\dag}) + \hbar g_{2}(a_2~ |b_1\rangle\langle b_2| + |b_2\rangle\langle
b_1|~ a_2^{\dag})~.
\end{equation}

The operator $|j\rangle\langle j|(j=a,b_1,b_2)$ gives the projection on to 
the state $|j\rangle$ with energy $\hbar{\Omega}_j$.
The operators $|a\rangle\langle b_1|$ and $|b_1\rangle\langle b_2|$ describe
the atomic transitions from the upper and lower levels to the middle level.
The operators $a_{\alpha}$ ($a_{\alpha}^{\dag}$) annihilate (create) a photon 
in the modes $\alpha = 1,2$ with resonance frequencies $\omega_{1} = {\Omega}_a
- {\Omega}_{b_{1}}$ and $\omega_{2} = {\Omega}_{b_{1}} - {\Omega}_{b_{2}}$ respectively. 
The first and second terms in the interaction operator $H_{AF}$ represents the
action of fields 1 and 2 of the cavity on the upper $(a \Leftrightarrow b_1)$
and the lower $(b_1 \Leftrightarrow b_2)$ transitions respectively.   
The parameters $g_{\alpha}$ are the corresponding atom-field coupling constants 
and $m$ is the atomic mass. The parameters $g_{\alpha}$ are dependent on z through 
the mode function of the cavity.

In the interaction picture, the Hamiltonian $(\ref{ham1})$ of the atom-field
system reads 
\begin{equation}
H_{I} = \frac{p_z^{2}}{2 m} + H_{AF}~.
\label{ham2}
\end{equation}
It is useful to expand the interaction Hamiltonian $H_{AF}$ in its diagonal basis.
The operator $H_{AF}$ has eigenstates $|\phi^{0}_{n_{1} + 1, n_{2} + 1}\rangle$,
$|\phi^{\pm}_{n_{1} + 1, n_{2} + 1}\rangle$ with eigenvalues $0, \pm \hbar 
\sqrt{g_1^{2}(z) (n_1 + 1) + g_2^{2}(z) (n_2 + 1)}$, respectively, where
\[
|\phi^{0}_{n_{1} + 1, n_{2} + 1}\rangle = \frac{g_2 \sqrt{n_2 + 1}}{\sqrt{g_1^{2}
(n_1 + 1) + g_2^{2} (n_2 + 1)}} |a,n_1,n_2\rangle~-~\frac{g_1 \sqrt{n_1 + 1}}
{\sqrt{g_1^{2} (n_1 + 1) + g_2^{2} (n_2 + 1)}} |b_2,n_1 + 1,n_2 + 1\rangle
\]

\begin{eqnarray}
|\phi^{\pm}_{n_{1} + 1, n_{2} + 1}\rangle &=& \frac{1}{\sqrt{2}} \left[ 
\frac{g_1 \sqrt{n_1 + 1}}{\sqrt{g_1^{2} (n_1 + 1) + g_2^{2} (n_2 + 1)}}
|a,n_1,n_2\rangle~ \pm ~ |b_1,n_1 + 1,n_2\rangle \right.\nonumber \\
&~~~& \left. + \frac{g_2 \sqrt{n_2 + 1}}{\sqrt{g_1^{2} (n_1 + 1) + g_2^{2} 
(n_2 + 1)}} |b_2,n_1 + 1,n_2 + 1 \rangle \right]
\end{eqnarray}
The interaction operator $H_{AF}$ and its eigenstates $|\phi^{0}_{n_{1} + 1, 
n_{2} + 1}\rangle$, $|\phi^{\pm}_{n_{1} + 1, n_{2} + 1}\rangle$  depend on the
position $z$ through the coupling strengths $g_1(z)$, $g_2(z)$. Thus, it is
in general difficult to carry out the time evolution of an atom-field state
governed by the Hamiltonian $(\ref{ham2})$ for the quantized motion of
atoms. So, for simplicity, we work with the mesa mode functions $g_{\alpha}(z) = 
g_{\alpha}~\theta(z) \theta(L - z)$ which represent the constant field modes in  
the cavity. In this case, the eigenstates of the interaction are independent 
of atomic position inside the cavity and the atomic motion sees free particle 
evolution in the dark state of interaction $|\phi^{0}_{n_{1} + 1,n_{2} + 1}\rangle$.
The effect of atom's interaction with the cavity on its external motion can be 
realized only in the dressed state $|\phi^{\pm}_{n_{1} + 1,n_{2} + 1}\rangle$ 
components of the initial atom-field state. We need to consider the initial 
atom-field state to be $|a,n_1,n_2\rangle$, i.e., the atom is in the excited state
and the cavity contains $(n_1,n_2)$ photons in the modes $(1,2)$ initially. This 
initial state can be expanded in the dressed state basis as
\begin{eqnarray}
|a,n_1,n_2\rangle &=& \left[\frac{g_1 \sqrt{(n_1 + 1)/2}}{\sqrt{g_1^{2}
(n_1 + 1) + g_2^{2} (n_2 + 1)}} \left(|\phi^{+}_{n_{1} + 1, n_{2} + 1}
\rangle  + |\phi^{-}_{n_{1} + 1, n_{2} + 1}\rangle \right) \right.
\nonumber \\ 
&~~~~~~~~~~~& \left. + \frac{g_2 \sqrt{n_2 + 1}}{\sqrt{g_1^{2} (n_1 + 1)
+ g_2^{2} (n_2 + 1)}} |\phi^{0}_{n_{1} + 1, n_{2} + 1}\rangle \right]~.
\label{expand}
\end{eqnarray}
The time evolution of this initial state can be found by expanding the
combined state of atom-cavity system as 
\begin{equation}
|\Psi(z,t)\rangle = \chi_{+}(z,t) |\phi^{+}_{n_{1} + 1, n_{2} + 1}
\rangle + \chi_{-}(z,t) |\phi^{-}_{n_{1} + 1, n_{2} + 1}\rangle +
\chi_{0} |\phi^{0}_{n_{1} + 1, n_{2} + 1}\rangle~,
\label{main1}
\end{equation}
then the time dependent Schrodinger equation becomes 
\begin{equation}
i \hbar \frac{\partial \chi_{\alpha}(z,t)}{\partial t} = h_{\alpha}
\chi_{\alpha}(z,t)~, ~~~~~\alpha = \pm,0.
\label{main2}
\end{equation}
Here, $h_{\pm} = p_z^{2}/2 m \pm \hbar \sqrt{g_1^{2} (n_1 + 1)
+ g_2^{2} (n_2 + 1)}$, $h_{0} = p_{z}^{2}/2 m $ are operators 
acting in the space of center of mass variables. Thus, the effect
of the cavity with fixed number of photons produces potential terms
in $h_{\alpha}$ corresponding to the dressed states $|\phi^{\pm}_
{n_{1} + 1, n_{2} + 1}\rangle$ as discussed in Ref. \cite{meyer}. 
The barrier and well potentials induced by the interaction for the
atomic motion in the states $|\phi^{\pm}_{n_{1} + 1, n_{2} + 1}\rangle$
are then displayed as in Fig. 2. It is also important to note that the
external motion of atom experiences free time evolution in the dark state
$|\phi^{0}_{n_{1} + 1, n_{2} + 1}\rangle$ for the mesa mode distribution   
of the cavity fields. Denoting the reflection and transmission amplitudes     
as $\rho^{\pm}_{n_1,n_2}$, $\tau^{\pm}_{n_1,n_2}$ for the potential
barrier-well problem of the dressed states $|\phi^{\pm}_{n_{1} + 1, 
n_{2} + 1}\rangle$, respectively, we have
\begin{equation}
{\rho}_{n_1,n_2}^{\pm} = i {\Delta}_{n_1,n_2}^{\pm} \sin(
k_{n_1,n_2}^{\pm} L) \exp(i k L) {\tau}_{n_1,n_2}^{\pm}~,
\end{equation}
\begin{equation}
{\tau}_{n_1,n_2}^{\pm} = \exp(-i k L) {\left [\cos(k_{n_1,n_2}^{\pm} L) - i
{\Sigma}_{n_1,n_2}^{\pm}\sin(k_{n_1,n_2}^{\pm} L) \right] }^{-1}~,
\label{taus}
\end{equation}
\begin{equation}
{\Delta}_{n_1,n_2}^{\pm} = \frac{1}{2}\left(\frac{k_{n_1,n_2}^{\pm}}
{k} - \frac{k}{k_{n_1,n_2}^{\pm}}\right)~,
\label{dspm}
\end{equation}
\[
{\Sigma}_{n_1,n_2}^{\pm} = \frac{1}{2}\left(\frac{k_{n_1,n_2}^{\pm}}
{k}+\frac{k}{k_{n_1,n_2}^{\pm}}\right)~,
\]
\begin{eqnarray}
k_{n_1,n_2}^{\pm} &=& \sqrt{\left(k^2 \mp \frac{2m}{\hbar}
\sqrt{g_{1}^{2}(n_{1}+1)+g_{2}^{2}(n_{2}+1)}\right)}
\label{kpm}~,   
\end{eqnarray}
where $\hbar k$ is the atomic c.m. momentum and $L$ is the length of the
cavity. It is to be noted that the strengths of the barrier - well
potentials (potential energy term in $k_{n_1,n_2}^{\pm}$) depend on the 
coupling constants $g_1$, $g_2$ as well as the occupation numbers $n_1$,
$n_2$ of the photons in the cavity.

We consider the initial wave packet of the moving free
atom to be $\psi(z,t) = \exp\left(-i p_z^{2} t/2 m \hbar \right)
\int dk A(k) e^{ikz} = \int dk A(k) e^{-i \left(\hbar k^2/2 m\right) t}
e^{i k z}$. The Fourier amplitudes $A(k)$ are adjusted such that the
peak of the incident wave packet ${|\psi(0,t)|}^2$ at the entry $(z = 0)$
of the cavity occurs at time $t = 0$. The combined state of the 
atom-cavity system at the initial time $t = 0$ is therefore,
\begin{equation}
|\Psi(z,0)\rangle = \psi(z,0) |a,n_1,n_2\rangle~.
\label{initial}
\end{equation}
The wave function of the atom-field system at time $t$ is found by
solving the Eqs. $(\ref{main1})$ and $(\ref{main2})$ subject to
the above initial condition :
\begin{eqnarray}
|\Psi(z,t)\rangle &=& \int dk A(k) e^{-i \left(\hbar k^2/2 m\right) t}
\left\{\left[R_{a,n_1,n_2}(k) e^{-i k z} \theta(-z) + T_{a,n_1,n_2}(k)
e^{i k z} \theta(z-L)\right] |a,n_1,n_2\rangle \right. \nonumber \\ 
&~~& + \left[R_{b_1,n_{1} + 1,n_2}(k) e^{-i k z} \theta(-z) + 
T_{b_1,n_{1} + 1,n_2}(k) e^{i k z} \theta(z-L)\right] |b_1,n_{1} + 1,n_2
\rangle \nonumber \\
&~~& \left. + \left[R_{b_2,n_{1} + 1,n_{2} + 1}(k) e^{-i k z} \theta(-z) +
T_{b_2,n_{1} + 1,n_{2} + 1}(k) e^{i k z} \theta(z-L)\right]
|b_2,n_{1} + 1,n_2 + 1\rangle \right\}~,
\end{eqnarray}
where
\begin{eqnarray}
R_{a,n_1,n_2} &=& \frac{g_1^{2} (n_1 + 1)}{2\left(g_1^{2} (n_1 + 1) + 
g_2^{2} (n_2 + 1)\right)} \left(\rho_{n_1,n_2}^{+} + \rho_{n_1,n_2}^{-}\right)
~,\nonumber \\
T_{a,n_1,n_2} &=& \frac{g_1^{2} (n_1 + 1)}{2 \left(g_1^{2} (n_1 + 1) +
g_2^{2} (n_2 + 1)\right)} \left(\tau_{n_1,n_2}^{+} + \tau_{n_1,n_2}^{-}\right)
+ \frac{g_2^{2} (n_2 + 1)}{\left(g_1^{2} (n_1 + 1) + g_2^{2} (n_2 + 1)\right)}~,
\label{prob1}
\end{eqnarray}
are the reflection and transmission amplitudes for the excited state of the atom 
with the initial $(n_1,n_2)$ photons remaining in the two cavity modes and 
\begin{eqnarray}
R_{b_1,n_{1}+1,n_2} &=& \frac{g_1 \sqrt{(n_1 + 1)}}{2\sqrt{\left(g_1^{2}(n_1 + 1) 
+ g_2^{2}(n_2 + 1)\right)}}\left(\rho_{n_1,n_2}^{+} - \rho_{n_1,n_2}^{-}\right)   
~,\nonumber \\
T_{b_1,n_{1}+1,n_2} &=& \frac{g_1 \sqrt{(n_1 + 1)}}{2\sqrt{\left(g_1^{2}(n_1 + 1) 
+ g_2^{2}(n_2 + 1)\right)}}\left(\tau_{n_1,n_2}^{+} - \tau_{n_1,n_2}^{-}\right)~,
\label{prob2}
\end{eqnarray}
are the probability amplitudes that the excited atom goes to the state $|b_1
\rangle$ and emits a photon in mode 1 while getting reflected and transmitted
respectively. Similarly, the excited atom is reflected or transmitted and emits
a photon in both the cavity modes while making a transition to the ground state 
$|b_2\rangle$ via the middle state $|b_1\rangle$ with probability amplitudes  
\begin{eqnarray}
R_{b_2,n_{1}+1,n_{2}+1} &=& \frac{g_1 g_2 \sqrt{(n_1 + 1)(n_2 + 1)}}{2 \left
(g_1^{2}(n_1 + 1) + g_2^{2}(n_2 + 1)\right)} \left(\rho_{n_1,n_2}^{+} + 
\rho_{n_1,n_2}^{-}\right)~, \nonumber \\
T_{b_2,n_{1}+1,n_{2}+1} &=& \frac{g_1 g_2 \sqrt{(n_1 + 1)(n_2 + 1)}}{2 \left
(g_1^{2}(n_1 + 1) + g_2^{2}(n_2 + 1)\right)} \left(\tau_{n_1,n_2}^{+} + 
\tau_{n_1,n_2}^{-}\right) - \frac{g_1 g_2 \sqrt{(n_1 + 1)(n_2 + 1)}}{\left
(g_1^{2}(n_1 + 1) + g_2^{2}(n_2 + 1)\right)}~.
\label{prob3}
\end{eqnarray}
It is clear from the above equations that the effects of the barrier and 
well potentials induced by the dressed states $|\phi^{\pm}_{n_1 + 1, n_2 + 1}
\rangle$ add coherently in either the reflection or transmission of the atom.  
The additive term to the barrier-well amplitudes in Eqs. $(\ref{prob1})$ and
$(\ref{prob3})$ comes from the contribution of the dark state $|\phi^{0}_{
n_1 + 1, n_2 + 1}\rangle$ in the initial state expansion Eq. $(\ref{expand})$.
Since the dark state is orthogonal to the middle state $|b_1\rangle$, it 
influences only the two-photon emissions and not the one-photon emissions
of the excited atom. An important feature here is that the two-photon 
transition can always be induced by the field when the one-photon transition
is forbidden. This can be seen by examining the probabilities for different 
states of the atom. When an initially excited atom is incident upon the cavity
containing $(n_1,n_2)$ photons in the two cavity modes $(1,2)$, respectively, 
then from Eqs. $(\ref{prob2})$ and $(\ref{prob3})$ the probability that the 
atom makes a one-photon transition to the state $|b_1\rangle$ with the 
emission of a photon in mode 1 is 
\begin{equation}
P_{n_1,n_2}(a \rightarrow b_1) = {|R_{b_1,n_{1}+1,n_2}|}^2 + {|T_{b_1,n_{1}+1,
n_2}|}^2~,
\label{gain1}
\end{equation}
and the probability that the atom makes a two-photon transition to the state
$|b_2\rangle$ with the emission of a photon in each of the modes 1 and 2 of the 
cavity is  
\begin{equation}
P_{n_1,n_2}(a \rightarrow b_2) = {|R_{b_2,n_{1}+1,n_{2}+1}|}^2 + {|T_{b_2,n_{1}+1
,n_{2}+1}|}^2~.
\label{gain2}
\end{equation}
When $g_2 = 0$, the lower transition $b_1 \rightarrow b_2$ is forbidden and hence 
the two-photon transition probability $P_{n_1,n_2}(a \rightarrow b_2)$ vanishes.
Then, the upper transition $a \rightarrow b_1$ behaves like a two-level atom 
interacting with the mode 1 of the cavity field. In Fig. 3, we compare the 
photon emission probabilities of excited, two-level $(g_2 \equiv 0)$ and 
three-level $(g_2 \neq 0)$ atoms when the cavity is initially in vacuum state. 
We scale the parameters in terms of a wave number $\kappa$ which is  
defined by the vacuum coupling energy $\hbar g_1 \equiv \hbar^2 \kappa^2/2 m$ of
the two-level atom. Note that the coupling strength $g_1 = 2 \pi \times 10$ MHz of 
a $^{85}$Rb atom corresponds to $155~\mu$m length of the cavity for the parameter
$\kappa L = 20000 \pi$. The temperature of the atom is of the order of $10^{-8}$ K
for the mean momentum $k/\kappa = 0.01$. For ultra-cold, incident atoms 
$(k/\kappa << 1)$, the graph in Fig. 3(a) shows that the one-photon emission probability 
exhibits resonances as a function of length of the cavity for both two-level and 
three-level atoms. In the three-level atom, the resonances of the one-photon emission
probability occur at values of $\kappa L$ different from those of two-level atom. 
The two-photon emission probability shows maxima and minima at the resonance positions  
of the one-photon transition. This behavior arises from the interference of the 
contributions coming from different dressed states in the initial state expansion
Eq. $(\ref{expand})$. The transmission probability ${|T_{b_2,n_{1}+1,n_{2}+1}|}^2$ 
obtained from Eq. $(\ref{prob3})$ has an interference term proportional to the phase 
of the amplitude $\left(\tau_{n_1,n_2}^{+} + \tau_{n_1,n_2}^{-}\right)$. This term
can be constructive or destructive which leads to the enhancement or reduction of 
the two-photon emission probability. From the graph, we also see that $P_{0,0}(a
\rightarrow b_2) \approx 0.32$ when $P_{0,0}(a \rightarrow b_1) \approx 0$. This
implies that the probability of two-photon emission is not the product of the
probabilities for single photon emission. This also suggests that both the field
modes of the bimodal cavity can be amplified together through the two-photon transition 
of an excited atom. This feature is absent in the case of two-mode micromaser pumped  
by ultra-cold, $\Lambda$-type three-level atoms \cite{arun}. In the $\Lambda$-scheme 
micromazer, the excited atom can make only one-photon transition to either of the
two ground levels. Therefore, both the cavity modes can not be populated sequentially
by the photon emission from the same atom. Next, we compare our results for ultra-cold
atoms with that of fast, incident atoms in the $\Xi$- type configuration. 
In the case of fast, incident atoms $(k/\kappa >> 1)$, both one- and two- photon
emission probabilities exhibit Rabi oscillations as a function of the length of the 
cavity as seen from Fig. 3(b). The Rabi frequency of oscillation for one-photon 
emission is twice that of the two-photon emission. These features resemble exactly  
the results in the usual Jaynes-Cummings (JC) model where one neglects the quantization  
of the atomic motion in atom-cavity interaction \cite{eberly}. Thus, neglecting the
kinetic energy operator in the Hamiltonian $(\ref{ham1})$, the time evolution of
the initial atom-field state $|a,n_1,n_2\rangle$ gives for the photon emission 
probabilities in the JC model:  

\[
P_{n_1,n_2}(a \rightarrow b_1) = \frac{g_1^2 (n_1 + 1)}{\Omega^2} \sin^2(\Omega \tau)~,
\]
\[
P_{n_1,n_2}(a \rightarrow b_2) = \frac{4 g_1^2 g_2^2 (n_1 + 1) (n_2 + 1)}{\Omega^4}
\sin^4(\Omega \tau/2)~,
\]
where $\Omega \equiv \sqrt{g_1^2 (n_1 + 1) + g_2^2 (n_2 + 1)}$ is the Rabi frequency 
and $\tau$ is the interaction time of the atom with the cavity. Note that when $\Omega 
\tau$ is an odd number multiple of $\pi$, the one-photon transition is forbidden and
the two-photon transition probability becomes maximum consistent with the results for
fast atoms in Fig. 3.
 
\section{mazer action in a bimodal cavity - basic master equation for the cavity
field}
In this section, we consider the mazer action of ultracold atoms in a bimodal 
cavity and derive the master equation for the cavity field assuming that excited 
atoms are pumped steadily into the cavity. We model the random pumping of the atoms  
into the cavity by a Poissonian process with an average rate of pumping $r$. The flux  
of the incident atoms is adjusted so that only one atom interacts with the cavity field  
at a time. We neglect the cavity field damping during the time an atom interacts with  
the field. Since the field in the cavity changes with the passage of each atom, we
need to know the time evolution of the atom-field state for a general initial state
of the cavity field. The wave function of the initial atom-field system is now
given by 
\begin{equation}
|\Psi(z,0)\rangle = \psi(z,0) \sum_{n_1,n_2} C_{n_1,n_2} |a,n_1,n_2\rangle.
\label{initials}
\end{equation}
Carrying out the time evolution for this initial state using Eqs. 
$(\ref{expand})$-$(\ref{main2})$, the state of atom-field system after 
the interaction is given by
\begin{eqnarray}
\mid \Psi(z,t)\rangle &=& \int dk A(k)e^{-i(\hbar k^2/2 m)t} \nonumber \\
& & \sum_{n_1,n_2 =0}^{\infty} \left[{\cal R}_{a,n_1,n_2}(k)e^{-ikz}\theta(-z)
\mid a,n_1,n_2\rangle +{\cal T}_{a,n_1,n_2}(k) e^{ikz}\theta(z-L)
\mid a,n_1,n_2\rangle \right. \nonumber \\
&+& {\cal R}_{b_1,n_{1}+1,n_2}(k) e^{-ikz}\theta(-z)
\mid b_1,n_{1}+1,n_2\rangle + {\cal T}_{b_1,n_{1}+1,n_2}(k)e^{ikz}
\theta(z-L) \label{general} \\
& &\mid b_1,n_{1}+1,n_2\rangle +{\cal R}_{b_2,n_{1}+1,n_{2}+1}(k)e^{-ikz}
\theta(-z)\mid b_2,n_{1}+1,n_{2}+1\rangle +{\cal T}_{b_2,n_{1}+1,n_{2}+1}(k)
\nonumber \\
& &\left.e^{ikz}\theta(z-L)\mid b_2,n_{1}+1,n_{2}+1\rangle \right]~, 
\nonumber
\end{eqnarray}
where
\begin{eqnarray}
{\cal R}_{a,n_1,n_2}(k) &=& C_{n_1,n_2} R_{a,n_1,n_2}(k)~, 
\label{lower} \nonumber \\
{\cal T}_{a,n_1,n_2}(k) &=& C_{n_1,n_2} T_{a,n_1,n_2}(k)~,
\end{eqnarray}
are the probability amplitudes for reflection (or) transmission of the atom
in the upper state $|a\rangle$ with $(n_1,n_2)$ photons in the cavity field
and similarly, the atom is reflected (or) transmitted when the atom-field
state is $|b_1,n_{1}+1,n_2\rangle$ (or) $|b_2,n_{1}+1,n_{2}+1\rangle$ with
amplitudes
\begin{eqnarray}
{\cal R}_{b_1,n_{1}+1,n_2}(k) &=& C_{n_1,n_2} R_{b_1,n_{1}+1,n_2}(k)~,
\nonumber\\
{\cal T}_{b_1,n_{1}+1,n_2}(k) &=& C_{n_1,n_2} T_{b_1,n_{1}+1,n_2}(k)~,
\label{middle} \nonumber\\
{\cal R}_{b_2,n_{1}+1,n_{2}+1}(k) &=& C_{n_1,n_2} R_{b_2,n_{1}+1,n_{2}+1}(k)~,\\
{\cal T}_{b_2,n_{1}+1,n_{2}+1}(k) &=& C_{n_1,n_2} T_{b_2,n_{1}+1,n_{2}+1}(k)~. 
\nonumber
\end{eqnarray}
The time evolution of the reduced density operator of the field in the 
interaction picture is then given in the coarse graining method \cite{lugiato}
to be
\begin{equation}
\dot{\rho}(t) = r \delta\rho(t) + L \rho(t)~,
\label{reduce}
\end{equation}
where $\delta\rho(t) = \rho(t) - \rho(0)$ is the change in the reduced density
operator of the field due to the passage of a single atom in the excited state.
This can be obtained by forming the atom-field density matrix using Eqs. 
$(\ref{initials})$ - $(\ref{middle})$ and tracing over external and internal 
degrees of freedom of the atom. Field damping and the effect of thermal photons
are described by the Liouville operator
\begin{eqnarray}
L\rho &=& \frac{1}{2} C_1(n_{b_{1}}+1)(2 a_1\rho a_{1}^{\dagger}
- a_{1}^{\dagger}a_1\rho - \rho a_{1}^{\dagger} a_1)\nonumber \\
&+& \frac{1}{2} C_1 n_{b_{1}}(2 a_{1}^{\dagger}\rho a_1 -
a_1 a_{1}^{\dagger} \rho - \rho a_1 a_{1}^{\dagger})\nonumber \\
&+& \frac{1}{2} C_2(n_{b_{2}}+1)(2 a_2\rho a_{2}^{\dagger}
- a_{2}^{\dagger}a_2\rho - \rho a_{2}^{\dagger} a_2) \label{damping} \\
&+& \frac{1}{2} C_2 n_{b_{2}}(2 a_{2}^{\dagger}\rho a_2 -
a_2 a_{2}^{\dagger} \rho - \rho a_2 a_{2}^{\dagger})\nonumber~.
\end{eqnarray}
Here $n_{b_{\alpha}}$ is the number of thermal photons in mode $\alpha$
and $C_{\alpha}$ is the damping rate of this mode. Using Eqs. 
$(\ref{reduce})$ and $(\ref{damping})$ we obtain the equation governing
the time evolution of density matrix elements, 

\begin{eqnarray}
\dot{\rho}(n_1,n_2;n_{1}^{\prime},n_{2}^{\prime}) &=&
r\left\{(R_{a,n_1,n_2} R_{a,{n_1}^{\prime},{n_2}^{\prime}}^{\star}
+ T_{a,n_1,n_2} T_{a,{n_1}^{\prime},{n_2}^{\prime}}^{\star} - 1)
\rho(n_1,n_2;n_{1}^{\prime},n_{2}^{\prime})\right.\nonumber \\
&+&\left.(R_{b_1,n_1,n_2} R_{b_1,{n_1}^{\prime},{n_2}^{\prime}}^{\star}+
T_{b_1,n_1,n_2} T_{b_1,{n_1}^{\prime},{n_2}^{\prime}}^{\star})
\rho(n_{1}-1,n_2;{n_1}^{\prime}-1,{n_2}^{\prime})\right.\nonumber \\
&+&\left.(R_{b_2,n_1,n_2} R_{b_2,{n_1}^{\prime},{n_2}^{\prime}}^{\star}+
T_{b_2,n_1,n_2} T_{b_2,{n_1}^{\prime},{n_2}^{\prime}}^{\star})
\rho(n_{1}-1,n_{2}-1;{n_1}^{\prime}-1,{n_2}^{\prime}-1)\right\}\nonumber \\
&+& \frac{1}{2} C_1(n_{b_1}+1)[2\sqrt{(n_{1}+1)(n_{1}^{\prime}+1)}
\rho(n_{1}+1,n_2;n_{1}^{\prime}+1,n_{2}^{\prime})\nonumber \nonumber\\
&-&(n_1+n_{1}^{\prime})\rho(n_{1},n_2;n_{1}^{\prime},n_{2}^{\prime})]\nonumber\\
&+& \frac{1}{2} C_1 n_{b_1}[2\sqrt{n_1 n_{1}^{\prime}}
\rho(n_{1}-1,n_2;n_{1}^{\prime}-1,n_{2}^{\prime}) \label{master}\\
&-&(n_1+n_{1}^{\prime}+2)\rho(n_1,n_2;n_{1}^{\prime},n_{2}^{\prime})]\nonumber\\
&+& \frac{1}{2} C_2(n_{b_2}+1)[2\sqrt{(n_{2}+1)(n_{2}^{\prime}+1)}
\rho(n_1,n_{2}+1;n_{1}^{\prime},n_{2}^{\prime}+1)\nonumber\\
&-&(n_2+n_{2}^{\prime})\rho(n_1,n_2;n_{1}^{\prime},n_{2}^{\prime})]\nonumber\\
&+&\frac{1}{2} C_2 n_{b_2}[2\sqrt{n_2 n_{2}^{\prime}}
\rho(n_1,n_{2}-1;n_{1}^{\prime},n_{2}^{\prime}-1) \nonumber\\
&-&(n_2+n_{2}^{\prime}+2)\rho(n_1,n_2;n_{1}^{\prime},n_{2}^{\prime})]~.\nonumber
\end{eqnarray}

The diagonal elements of the density matrix $P(n_1,n_2) = \rho(n_1,n_2;n_1,n_2)$
which gives the joint distribution of photons in the two cavity modes, obeys
the following equation:  
\begin{eqnarray}
\dot{P}(n_1,n_2) &=& - G_{b_1,n_1,n_2} P(n_1,n_2) + G_{b_1,n_1 - 1,n_2}
P(n_{1}-1,n_2) \nonumber \\
&-& G_{b_2,n_1,n_2} P(n_1,n_2) + G_{b_2,n_1 - 1,n_2 - 1} P(n_{1}-1,n_{2}-1) 
\nonumber \\ 
&+& C_1 (n_{b_1}+1)\left[(n_1+1)P(n_{1}+1,n_2)- n_1 P(n_1,n_2)\right]
\label{rate} \\
&+&C_1 n_{b_1}\left[n_1 P(n_{1}-1,n_2)-(n_{1}+1) P(n_1,n_2)\right]\nonumber \\
&+& C_2 (n_{b_2}+1)\left[(n_2+1)P(n_1,n_{2}+1)- n_2 P(n_1,n_2)\right]\nonumber
 \\
&+&C_2 n_{b_2} \left[n_2 P(n_1,n_{2}-1)-(n_{2}+1) P(n_1,n_2)\right]~,\nonumber
\end{eqnarray}
where $G_{b_1,n_1,n_2} = r P_{n_1,n_2}(a \rightarrow b_1)$ and $G_{b_2,n_1,n_2}  
= r P_{n_1,n_2}(a \rightarrow b_2)$ are the gain coefficients for the atomic 
transitions with $P_{n_1,n_2}(a \rightarrow b_1)$ and $P_{n_1,n_2}(a \rightarrow
b_2)$ as defined in Eqs. $(\ref{gain1})$ and $(\ref{gain2})$. This is the master
equation for the two-mode micromaser describing the time evolution of photon 
distribution in the cavity. This equation behaves similar to a rate equation
for the probability and a simple physical meaning can be given to each term on 
the right hand side in terms of inflow and outflow of probabilities. The first
and second terms in the equation gives the effect of one-photon transitions while
the third and fourth terms correspond to the two-photon transitions of the
excited atoms.  
 
\section{photon statistics of the mazer field}
The steady state distribution of photons obeys the equation
\begin{equation}
\dot{P}(n_1,n_2) = 0
\label{steady}
\end{equation}
In the limit $g_2 \rightarrow 0$, the two-photon transition probability 
in Eq. $(\ref{gain2})$ tends to zero and therefore we can neglect 
the third and fourth terms containing $G_{b_2,n_1,n_2}$ in the master Eq.
$(\ref{rate})$. In this case, the upper transitions $a \rightarrow b_1$ behave
like two-level atoms interacting with the mode 1 of the cavity. The lower 
transitions $b_1 \rightarrow b_2$ and hence the two-photon transitions    
$a \rightarrow b_1 \rightarrow b_2$ are forbidden in the interaction.  
The steady state solution of the Eq. $(\ref{steady})$ can then be obtained
in analytical form by using the principle of detailed balance as discussed
by Meyer et al \cite{meyer}. The photon statistics in this {\bf two-level problem 
is a mixture of thermal and shifted thermal distributions}. When $g_2 \neq 0$, both 
the one- and two-photon effects of atomic transitions contribute in building up the
cavity field. The two-photon terms (third and fourth terms) in the master equation
have no counterpart in the decay terms and therefore the equation is {\bf not solvable
analytically} for the steady state distribution by the principle of {\bf detailed balance}
adopted in all the previous works on micromasers. In this general case, we integrate 
the master equation $(\ref{rate})$ numerically using fourth order Runge kutta method
to get the steady state solution. We {\bf do not} use any decorrelation approximation.
The photon distribution $P_1(n)$ and $P_2(n)$ in the cavity modes 1 and 2 are obtained
respectively using 
\begin{equation}
P_1(n) = \sum_{l=0}^{\infty} P(n,l),~~~~~~~P_2(n) = \sum_{l=0}^{\infty} P(l,n)~.
\label{prob}
\end{equation}
The normalized variances of photon distribution in the two cavity modes 
are defined by
\begin{equation}
\sigma_{\alpha}^2 = \frac{\langle n_{\alpha}^2 \rangle - {\langle 
n_{\alpha} \rangle}^2}{\langle n_{\alpha} \rangle}~,~~~~~~\alpha = 1,2~.
\end{equation}
In Fig. 4, we present the numerical results of the photon distribution
in steady state for ultra-cold, incident atoms $(k/\kappa << 1)$ by assuming
equal parameters for the decay terms $C_1 = C_2 = C$, $n_{b_1} = n_{b_2} 
= n_b$. The graph shows that for $g_1 < g_2$, the photon statistics in mode 1 
is super-Poissonian $(\sigma_{1}^2 > 1)$ while that of mode 2 is sub-Poissonian 
$(\sigma_{2}^2 < 1)$. The photon distribution for $g_1 > g_2$ is identical to 
that of $g_1 < g_2$ except that modes 1 and 2 are interchanged. When $g_1 = g_2$,
each mode of the cavity field exhibits Poissonian statistics $(\sigma_{\alpha}^2 
\approx 1)$ of mean $r/2C -1$. To understand these numerical results, we now 
approximate the master equation $(\ref{rate})$ by dropping the first and second terms 
corresponding to one-photon transitions. In fact, for the parameters of Fig. 4,
the barrier and well amplitudes are $\rho_{n_1,n_2}^{\pm} \approx -1$,
$\tau_{n_1,n_2}^{\pm} \approx 0$ for wide range of $n_1$ and $n_2$ values.
Therefore, the one-photon transition probability $P_{n_1,n_2}(a \rightarrow b_1)$ 
in Eq. $(\ref{gain1})$ is approximately zero. The two-photon emission probability
$P_{n_1,n_2}(a \rightarrow b_2)$ in Eq. $(\ref{gain2})$ can then be approximated
to be $2 g_1^2 g_2^2 (n_1 + 1) (n_2 + 1)/{(g_1^2 (n_1 + 1) + g_2^2 (n_2 + 1))}^2 $.
Note that this approximation is also consistent with the results for ultra-cold atoms
in Fig. 3. With these substitutions for the photon    
emission probabilities in gain coefficients, numerical integration of the master 
Eq. $(\ref{rate})$ again gives the same results. Moreover, in the absence of 
one-photon terms, the master equation is symmetric with respect to the labels 
1 and 2 of the two cavity modes. Thus, the Poissonian distribution of photons in  
each cavity mode is purely the effect of two-photon transitions of the excited atoms.
It should be emphasized that the transmission of atoms occurs only in the dark 
eigenstate component of the interaction during the field buildup in the cavity. The  
atoms interacting with the barrier-well component of the dressed states get reflected
always. But both the reflected and the transmitted atoms {\bf have equal probability} 
of two-photon emissions into the cavity. Since the one-photon emissions are forbidden 
in the interaction for ultracold atoms, the photon distribution of both the cavity
modes peaks around the same photon number in Fig. 4. In general, the photon 
distribution for mode 1 peaks at a higher photon number when compared with that of
mode 2 . This is because the mode 1 of the cavity can be populated by both one- and
two-photon emissions while the mode 2 can be populated by only two-photon emissions
from the incident atoms. We have found this behavior of steady state photon
distribution in the case of fast, incident atoms. In Fig. 5, we display the steady
state photon distribution when the micromaser is pumped by fast, incident atoms
$(k/\kappa >> 1)$ for the same parameters of Fig. 4(a). The graph shows that the 
field 1 is amplified more than the field 2 by the stimulated, photon emissions from 
the incident atoms. It is important to note that the field 2 has an influence on the
field 1 in the cavity through the interaction with the atoms. The effects of field 2
such as gain enhancement or gain reduction on field 1 have been already discussed in
a different context viz in a two-beam laser operating on cascade three-level atoms
\cite{zhu}. Next, we show the effect of thermal photons in the cavity on the steady 
state photon distribution in Fig. 6. For comparison, we have also plotted the photon 
distribution for the two-level problem $(g_2 \equiv 0)$. The length $\kappa L$ of 
the cavity is chosen to be at a resonance of the one-photon emission probability
for the initial, excited state of the two-level atom and three photons in mode 1 of
the cavity. The photon distribution in the two-level problem then looks similar to 
a mixture of thermal and shifted thermal distributions as discussed by Meyer et al 
\cite{meyer}. For three-level atoms, comparison with Fig. 4(a) shows that the photon 
distribution broadens because of the presence of thermal photons $(n_b \neq 0)$ even
though qualitative features are very similar. In particular, the two-photon 
emissions from the pumping atoms are still the dominant contribution to the steady 
state field for the chosen length of the cavity. The Poissonian-like statistics of
photons in the micromaser cavity pumped by cold atoms, resembles closely the
behavior of a laser operating at far above threshold \cite{book}. Finally, we
note that the two-photon effects are dominant over the one-photon transitions
only in the limit $(k/\kappa << 1)$ of ultra-slow motion of the incident atoms.
The competition of the one-photon with the two-photon processes becomes stronger
even for energies of the incident atoms $(k/\kappa \approx 1)$ close to the vacuum
coupling energy. In Fig. 7, we plot the photon distribution in the cavity for  
the mean momentum $k/\kappa = 1.1$ of the incident atoms. The inset of the Fig. 7 
shows the photon emission probabilities of an excited atom for an initial, vacuum 
state of the cavity field. The graph shows that the one-photon effects participate
in the field build up of the cavity and these lead to unequal, average number of
photons in the two modes of the mazer field.  

\section{summary}
We considered the interaction of mono-energetic beam of ultra-cold, cascade three-
level atoms in the excited state with a bimodal cavity. The atom-field interaction
is equivalent to reflection and transmission of the atoms through the potentials
induced by the dressed states. There is also a reflection-less transmission of the
atoms in the dark eigenstate of atom-field interaction. We find that two-photon
transition can always be induced in an excited atom when its one-photon transition is
forbidden. In general, the two-photon emissions from the excited atoms dominate over
the one-photon emissions in building up the steady state field of the cavity. The 
steady state photon distribution in each mode exhibits sub- and super- Poissonian
behaviors depending on the strengths of the atom-field couplings. The photon
distribution approaches the Poissonian statistics when the atom-field coupling
strengths of the two modes are equal. We have also obtained the photon distribution
in steady state when the micromaser is pumped by fast, moving atoms instead of
ultracold atoms. In this case, both one- and two- photon emissions from the 
incident atoms contribute in building up the steady field of the cavity.

\newpage
\vspace*{2.5 cm}
\begin{figure}[h]
\hspace*{2.8 cm}
\epsfxsize 3.2in
\epsfysize 2.2in
\epsfbox{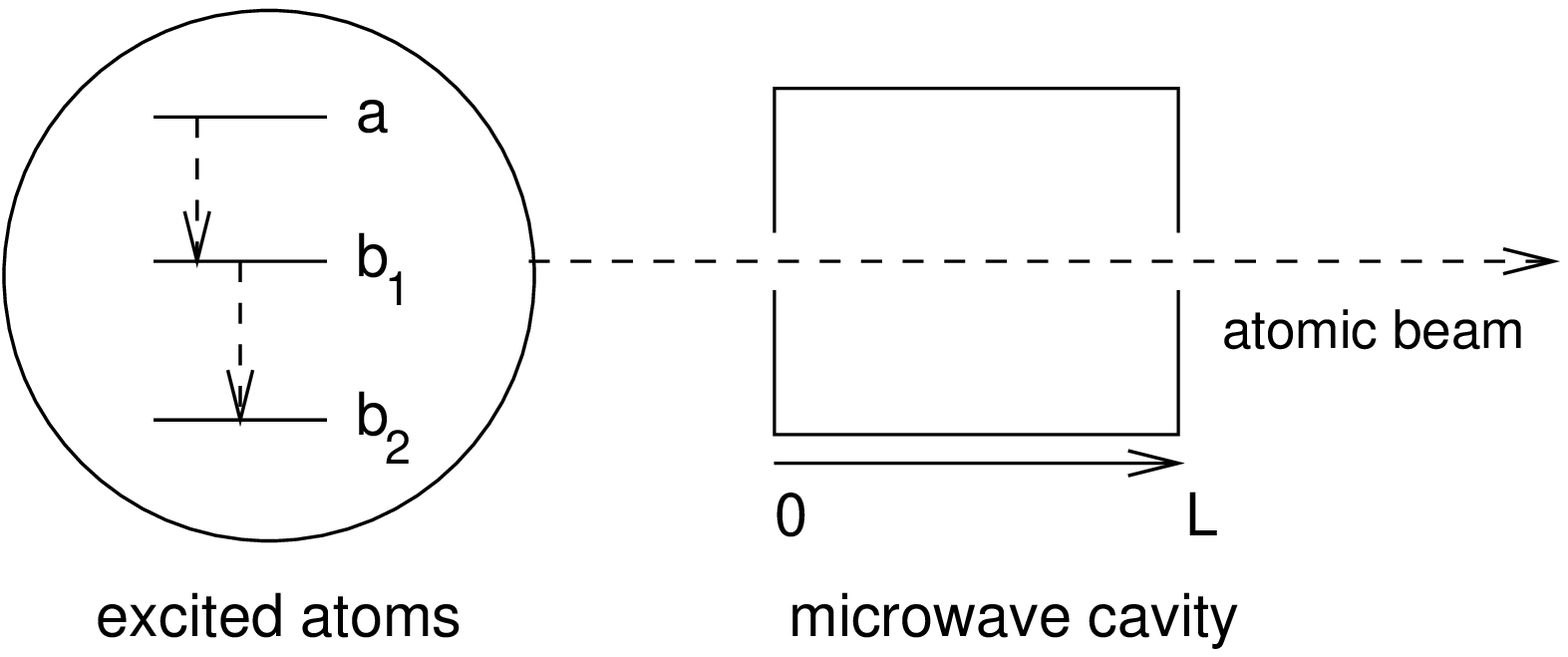}
\end{figure}
\vspace*{-2.3 cm}
FIG. 1.~~The scheme of the two-mode micromaser cavity pumped by $\Xi$-type
three-level atoms in the excited state.

\vspace*{2.7 cm}
\begin{figure}[h]
\hspace*{2.8 cm}
\epsfxsize 3.6in
\epsfysize 2.4in
\epsfbox{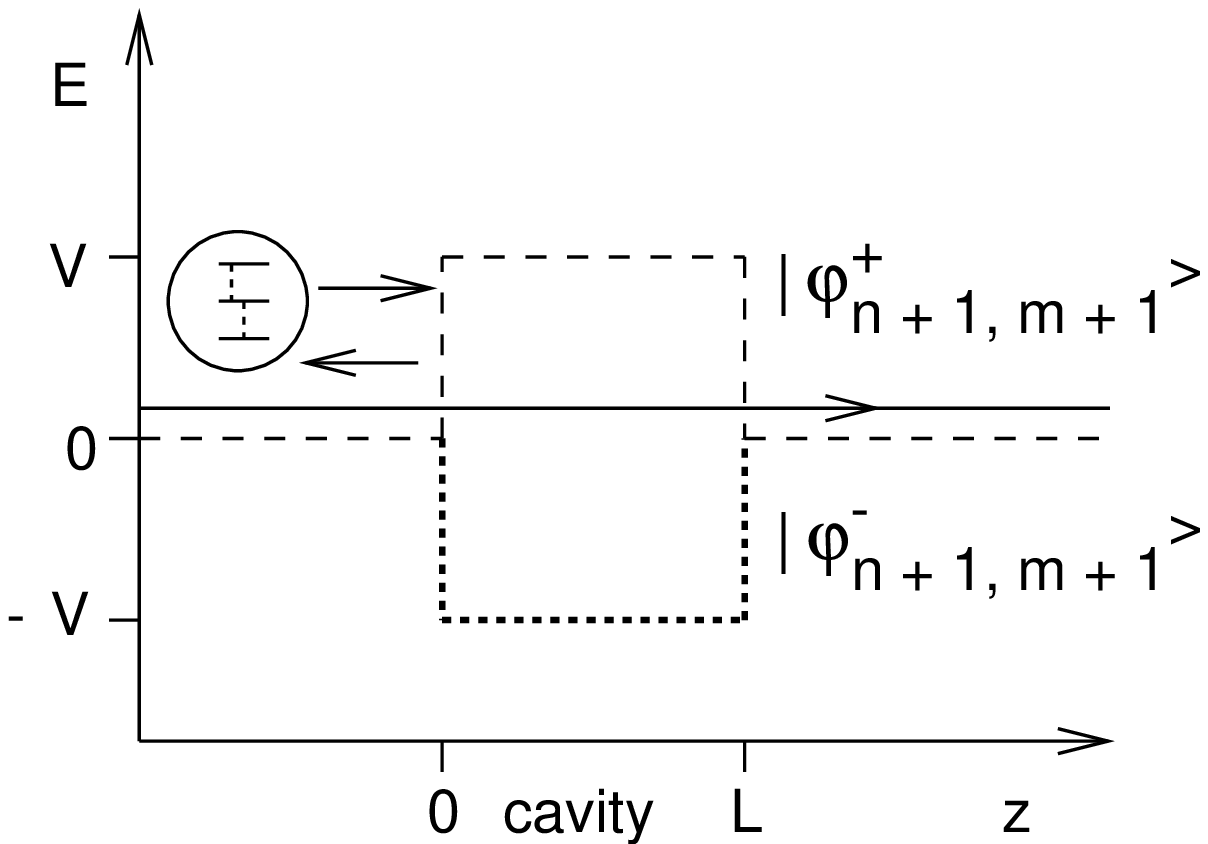}
\end{figure}
\vspace*{-0.1 cm}
FIG. 2.~~Schematic representation of the energy E of the excited atoms
incident upon a two-mode micromaser cavity with $(n,m)$ photons. The 
atom-field interaction creates a barrier (dashed) and well (dotted) 
potentials with a potential energy $V = \hbar \sqrt{g_1^{2} (n + 1) +
g_2^{2} (m + 1)}$ in the dressed states $|\phi_{n+1,m+1}^{\pm}\rangle$.
The scattering from these cavity induced potentials leads to reflection
or transmission of the atoms through the cavity. The interaction also induces 
a reflection-less transmission of the atoms in the dark state $|\phi_{n+1,m+1}^{0}
\rangle$. However, the reflection or transmission of the atoms can occur only
in either of the three states $|a,n,m\rangle$, $|b_1,n+1,m\rangle$, and
$|b_2,n+1,m+1\rangle$.

\newpage
\vspace*{2 cm}
\begin{figure}[h]
\hspace*{1.3 cm}
\epsfxsize=350pt
\epsfbox{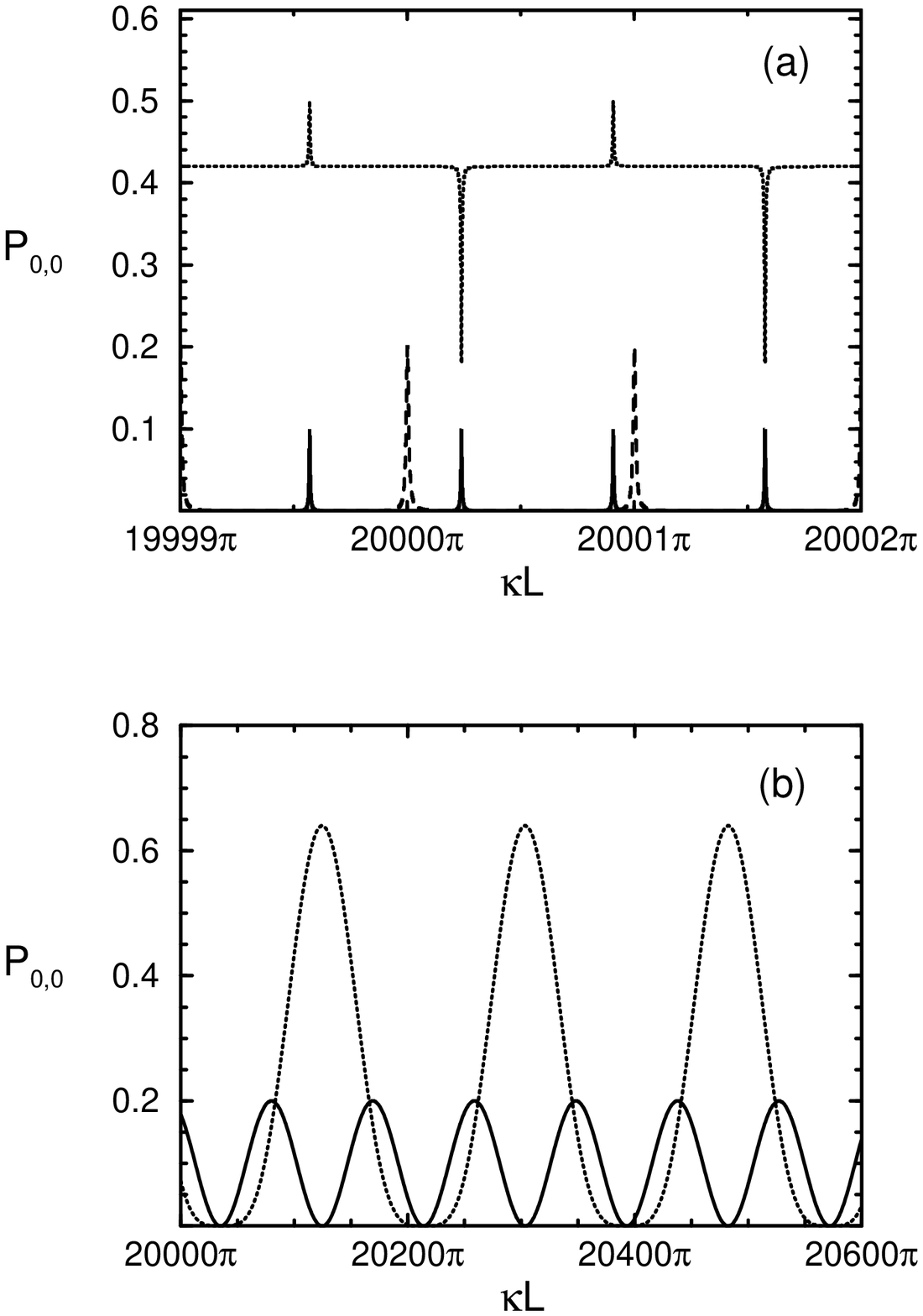}
\end{figure}
\vspace*{-2.1 cm}
FIG. 3.~~The probabilities of $a \rightarrow b_1$ (solid curve) 
and $a \rightarrow b_1 \rightarrow b_2$ (dotted curve) transitions of 
an excited atom as a function of the length $\kappa L$ of the cavity.
The cavity is initially in vacuum state and the parameters used are  
$g_2/g_1 = 2$, $k/\kappa = 0.01$ [(a)], $k/\kappa = 100$ [(b)]. The 
dashed curve in Fig. 3(a) represents the photon emission probability of
an excited two-level atom resonant with the upper transition when $g_2 = 0$.
Actual values of the dashed curve are 2.5 times those shown. For clarity, 
the dotted curve in Fig. 3(a) has been displaced by 0.1 units along
the Y axis.
 
\newpage
\vspace*{2 cm}
\begin{figure}[h]
\hspace*{1.3 cm}
\epsfxsize=350pt
\epsfbox{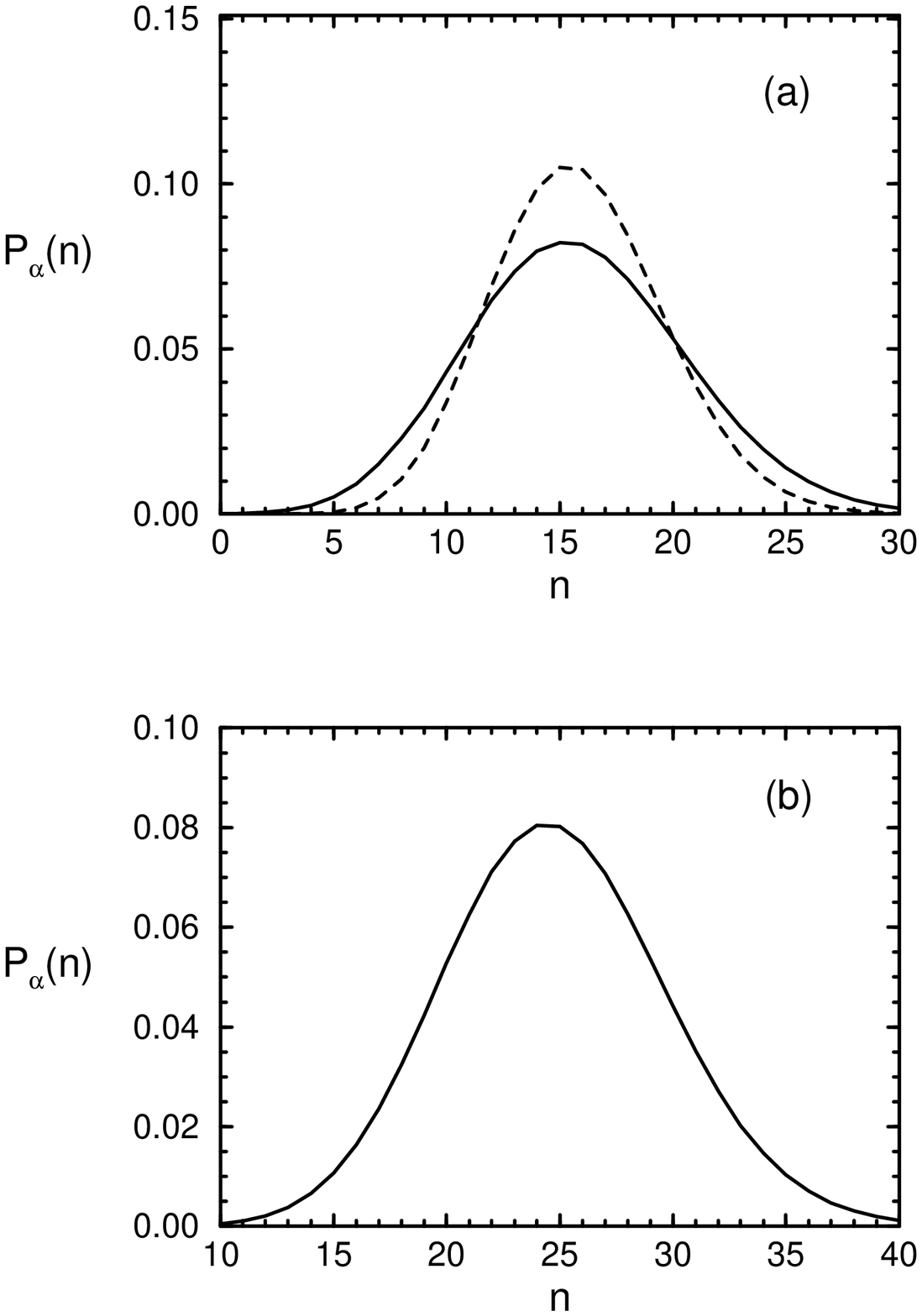}
\end{figure}
\vspace*{-2.1 cm}
FIG. 4.~~The steady state distribution of photons in mode 1 (solid curve) and 
mode 2 (dashed curve). The parameters used are $C_1 = C_2 = C$, $n_{b_1} =
n_{b_2} = n_{b}$, $r/C = 50$, $n_b = 0$, $\kappa L = 20000 \pi$, $k/\kappa = 
0.01$ and (a) $g_2/g_1 = 2$, (b) $g_2/g_1 = 1$. In the case of $g_2 = g_1$,
the dashed curve is hardly distinguishable from the solid curve. The photon
statistics for the parameter $g_2/g_1 = 0.5$ is approximately similar to that
of $g_2/g_1 = 2$ except that the solid (dashed) curve corresponds to the
photon distribution in mode 2 (1).
 
\newpage
\vspace*{2 cm}
\begin{figure}[h]
\hspace*{1.5 cm}
\epsfxsize=320pt
\epsfbox{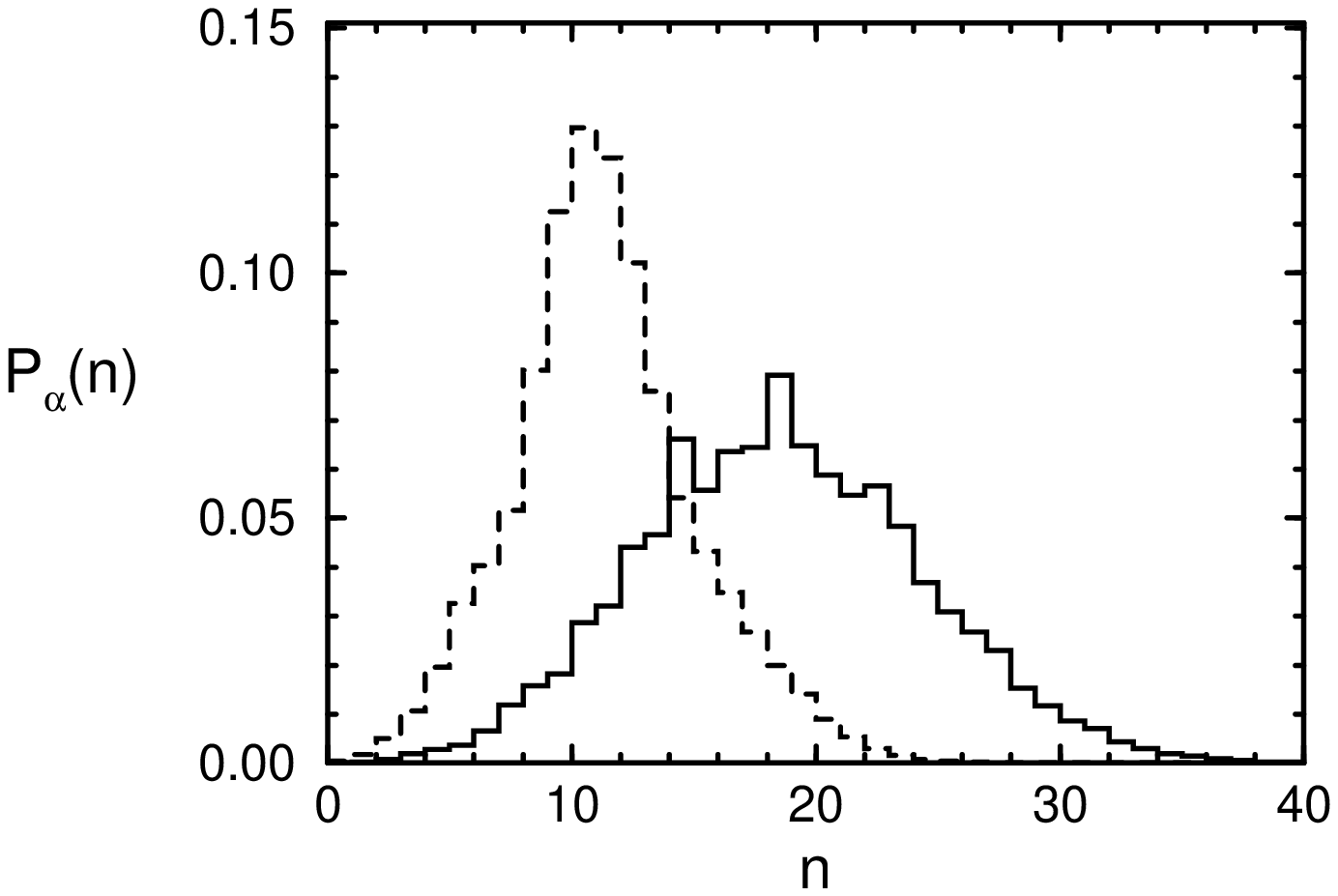}
\end{figure}
\vspace*{-1 cm}
FIG. 5.~~The steady state distribution of photons in mode 1 (solid) and
mode 2 (dashed) for the same parameters of Fig. 4(a) with $k/\kappa = 100$. 

\vspace*{1.4 cm}
\begin{figure}[h]
\hspace*{1.5 cm}
\epsfxsize=320pt
\epsfbox{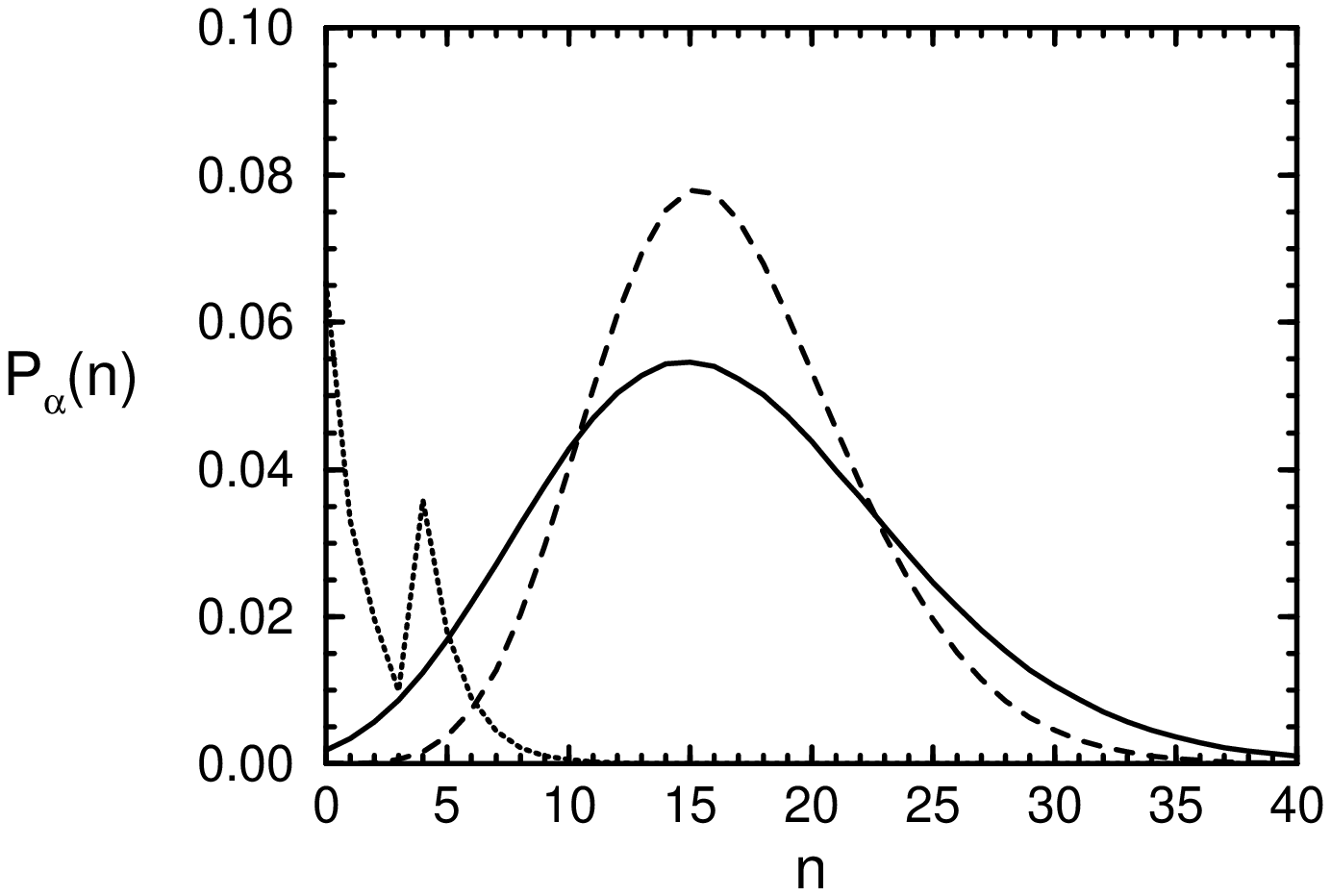}
\end{figure}
\vspace*{-1 cm}
FIG. 6.~~The steady state distribution of photons in mode 1 (solid curve) and
mode 2 (dashed curve). The parameters for the calculation are $C_1 = C_2 = C$, 
$n_{b_1} = n_{b_2} = n_{b}$, $r/C = 50$, $k/\kappa = 0.01$, $g_2/g_1 = 2$, 
$\kappa L = 40000 \pi/\sqrt[4]{4}$ and $n_b = 1$. The dotted curve represents
the photon distribution in mode 1 for the two-level problem when $g_2 = 0$.
Actual values of the dotted curve are 5 times those shown.    

\newpage
\begin{figure}[h]
\hspace*{2.6 cm}
\epsfxsize=270pt
\epsfbox{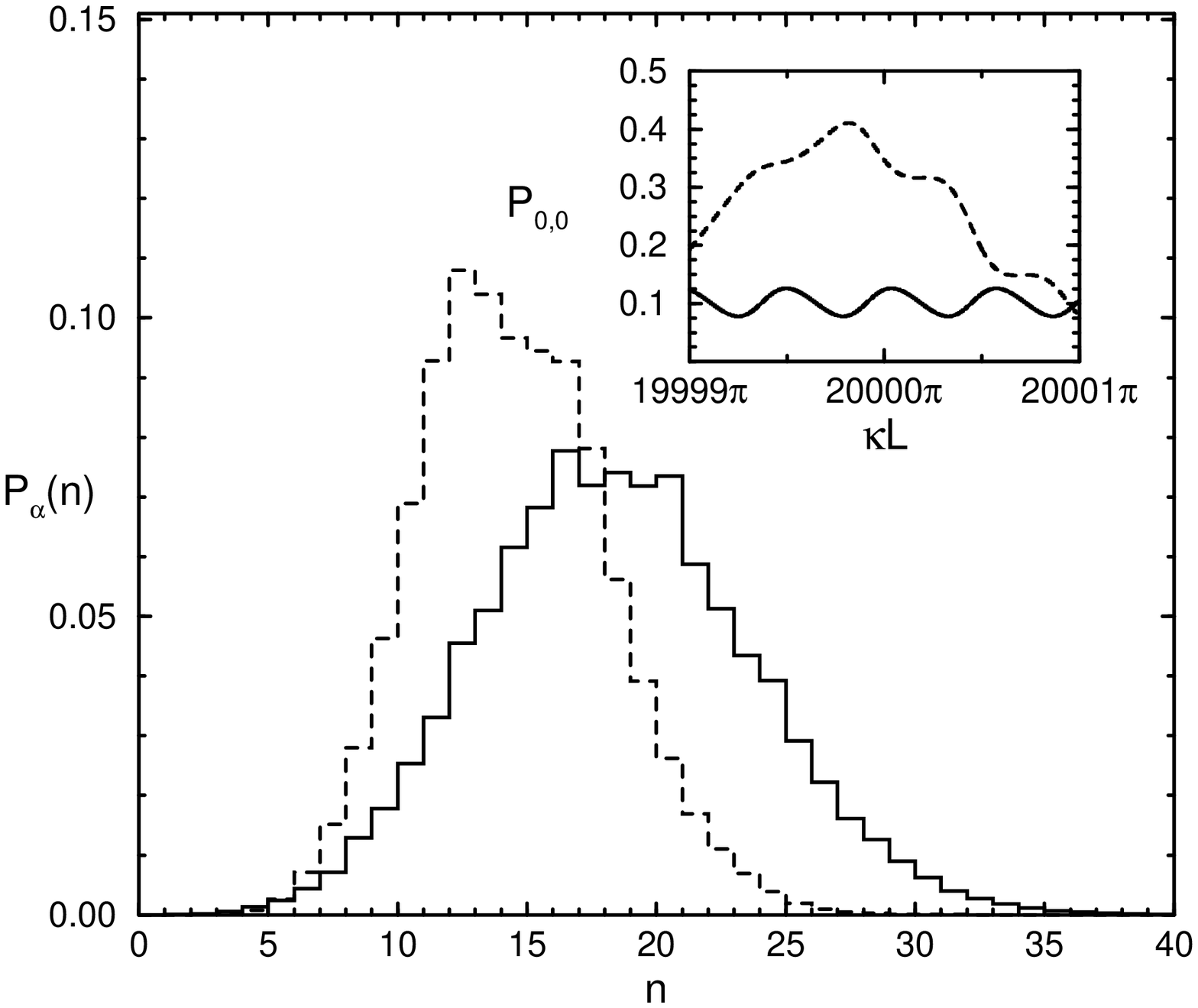}
\end{figure}
\vspace*{-4.0 cm}
FIG. 7.~~The steady state distribution of photons in mode 1 (solid) and mode 2
(dashed) for the same parameters of Fig. 4(a) with $k/\kappa = 1.1$. The
solid (dashed) curves in the inset represents the one-photon (two-photon)
transition probabilities of an excited atom for an initial vacuum field in
the cavity. The parameters used for the inset are $g_2/g_1 = 2$ and $k/\kappa = 
1.1.$ 

\end{document}